\documentclass[amssymb,amsmath,aps,groupedaddress,reprint,superscriptaddress,twocolumn,showpacs]{revtex4}

\usepackage{graphicx}
\usepackage{dcolumn}
\usepackage{bm}
\usepackage{amsmath}

\begin{document}

\title{Efficient nonlinear generation of THz plasmons in graphene and topological insulators}
\author{Xianghan Yao}
\affiliation{Department of Physics and Astronomy, Texas A\&M
University, College Station, TX, 77843 USA}
\author{Mikhail Tokman}
\affiliation{Institute of Applied Physics, Russian Academy of Sciences}
\author{Alexey Belyanin}
\affiliation{Department of Physics and Astronomy, Texas A\&M
University, College Station, TX, 77843 USA}

\date{\today}

\begin{abstract}

Surface plasmons in graphene may provide an attractive alternative to noble-metal plasmons due to their tighter confinement, peculiar dispersion, and longer propagation distance. We present theoretical studies of the nonlinear difference frequency generation of terahertz surface plasmon modes supported by two-dimensional layers of massless Dirac electrons, which includes graphene and surface states in topological insulators. Our results demonstrate strong enhancement of the DFG efficiency near the plasmon resonance and the feasibility of phase-matched nonlinear generation of plasmons over a broad range of frequencies.

\end{abstract}

\pacs{81.05.ue, 42.65.-k}

\maketitle

Graphene exhibits many interesting electronic properties because of its chiral symmetry and gapless linear spectrum of free carriers near the Dirac point. In recent years, graphene has also been recognized as a promising broadband optoelectronic material in the infrared (IR) and terahertz (THz) region, especially when utilizing a surface plasmon resonance  \cite{ju2011,grigorenko2012,tredicucci}.
Surface plasmon is a collective mode of coupled charge-density and field oscillations at an interface between a free-carrier system and a dielectric or vacuum. Surface plasmons guided by graphene are expected to have low losses and be highly tunable by gating and doping, making graphene an attractive alternative to metal plasmonics. Surface states in certain topological insulators (TIs) have a massless Dirac-cone electron dispersion at low energies with a slope similar to that in graphene. They provide a potentially even more interesting host medium for surface plasmons due to lower scattering rates of two-dimensional (2D) surface
electrons that are topologically protected from scattering on non-magnetic impurities \cite{hasan}. In particular, Bi$_2$Se$_3$  has a large bulk band gap of about 0.3 eV, suitable for THz and mid-infrared plasmonics, and a tunable Fermi level which can be put within the bulk gap \cite{zhang}. The combination of highly efficient light-matter interaction, relatively long propagation distances, and tight confinement of surface plasmons in graphene and TIs promises interesting applications  including compact room-temperature THz sources for imaging, spectroscopy and telecommunications; integrated photonic circuits; THz modulation of telecom signals, and compact THz sensors. Furthermore, optical  methods \cite{shaf} may provide a better access to characterization and manipulation of massless fermion states than transport measurements that are affected by conductivity in the bulk.

Nonlinear optics of massless Dirac fermions has received little attention so far, especially in the THz range where many basic devices and components are lacking. Here we show that the difference frequency generation (DFG) in 2D layers of massless Dirac electrons, e.g. graphene and TIs, is an efficient and controllable way of generating surface plasmons over a broad range of frequencies.  Second-order nonlinear processes such as DFG are usually assumed to be forbidden in an isotropic medium \cite{boyd} such as the plane of a graphene layer. However, the second-order susceptibility $\chi^{(2)}$ becomes non-zero when its spatial dispersion (dependence on photon wave vectors) is taken into account.  In our case the anisotropy is induced by the in-plane wave vectors of obliquely incident electromagnetic waves. This effect is well-known for the DFG of plasma waves in a bulk isotropic classical plasma \cite{plasma}. Another second-order nonlinear process, second-harmonic generation in graphene has been theoretically studied in \cite{mikhailovSHG,glazov}. Of course second-order processes are also non-zero for out-of-plane excitations due to anisotropy between in-plane and out-of-plane directions, which is a property of any surface. Here we consider only in-plane excitations which yield a much stronger nonlinear effect. We find that the DFG of surface plasmons at the beat frequency of two  obliquely incident or in-plane propagating infrared beams shows a surprisingly high efficiency over a broad range of frequencies and is widely tunable by varying an angle of incidence, gating, or doping.

First we review the dispersion of surface plasmons in graphene and TIs, which has been studied before a number of times in various approximations; see e.g. \cite{jablan, efimkin,TI_plasmon}. Assuming the monolayer of massless Dirac electrons in the $xy$-plane, z-axis pointing up as in Fig. 1(a), and wave propagation in the $x$-direction, we calculate the linear response and the dispersion for TM surface plasmon modes \cite{suppl}. We include only the intraband transitions assuming that the electrons are degenerate and the Fermi energy is higher than the plasmon energy: $E_F > \hbar \omega/2$. This assumption can be easily dropped if needed. The resulting 2D linear electric susceptibility is given by $\chi_{xx} = \chi_{yy} = \tilde{\chi}_{\omega,q}$ and $\chi_{xy} = \chi_{yx} = 0$, where
\begin{equation}
\label{chi1}
\tilde{\chi}_{\omega,q} = \frac{g}{(2\pi)^2}\frac{e^2E_F}{\hbar^2\omega}\int^{2\pi}_{0}d\phi\frac{\cos^2{\phi}}{-i\gamma + \upsilon_Fq\cos{\phi}-\omega}.
\end{equation}
Here $\omega$ and $q$ are frequency and $x$-component of the wave vector in a monochromatic wave, $\phi$ is the angle between the electron momentum and $x$-axis; $\gamma$ is the scattering rate which greatly depends on the material and substrate quality. 
The degeneracy factor $g$ is equal to 4 for graphene and 1 for one surface in Bi$_2$Se$_3$. The slope of the linear electron dispersion at low energies, $\upsilon_F$ is $\sim 10^8$ cm/s in graphene and about two times lower in Bi$_2$Se$_3$ \cite{zhang}.  The integral in Eq. (\ref{chi1}) can be evaluated analytically.

For a TM-mode with non-zero $E_x$, $E_z$, and $B_y$ field components, using Maxwell's equations and standard boundary conditions \cite{suppl}, we can obtain the dispersion relation for surface modes guided by monolayer graphene, placed on an interface between two dielectric media with dielectric constants $\epsilon_1$ and  $\epsilon_2$, as shown in the inset to Fig. 1(a):
\begin{equation}
\label{mono}
D(\omega,q)=4\pi\tilde{\chi}_{\omega,q}+\frac{\epsilon_1}{p_1}+\frac{\epsilon_2}{p_2}=0,
\end{equation}
with $p_{1,2}=\sqrt{q^2-\epsilon_{1,2}\frac{\omega^2}{c^2}}$.

\begin{figure}[htb]
\centerline{
\includegraphics[width=9cm]{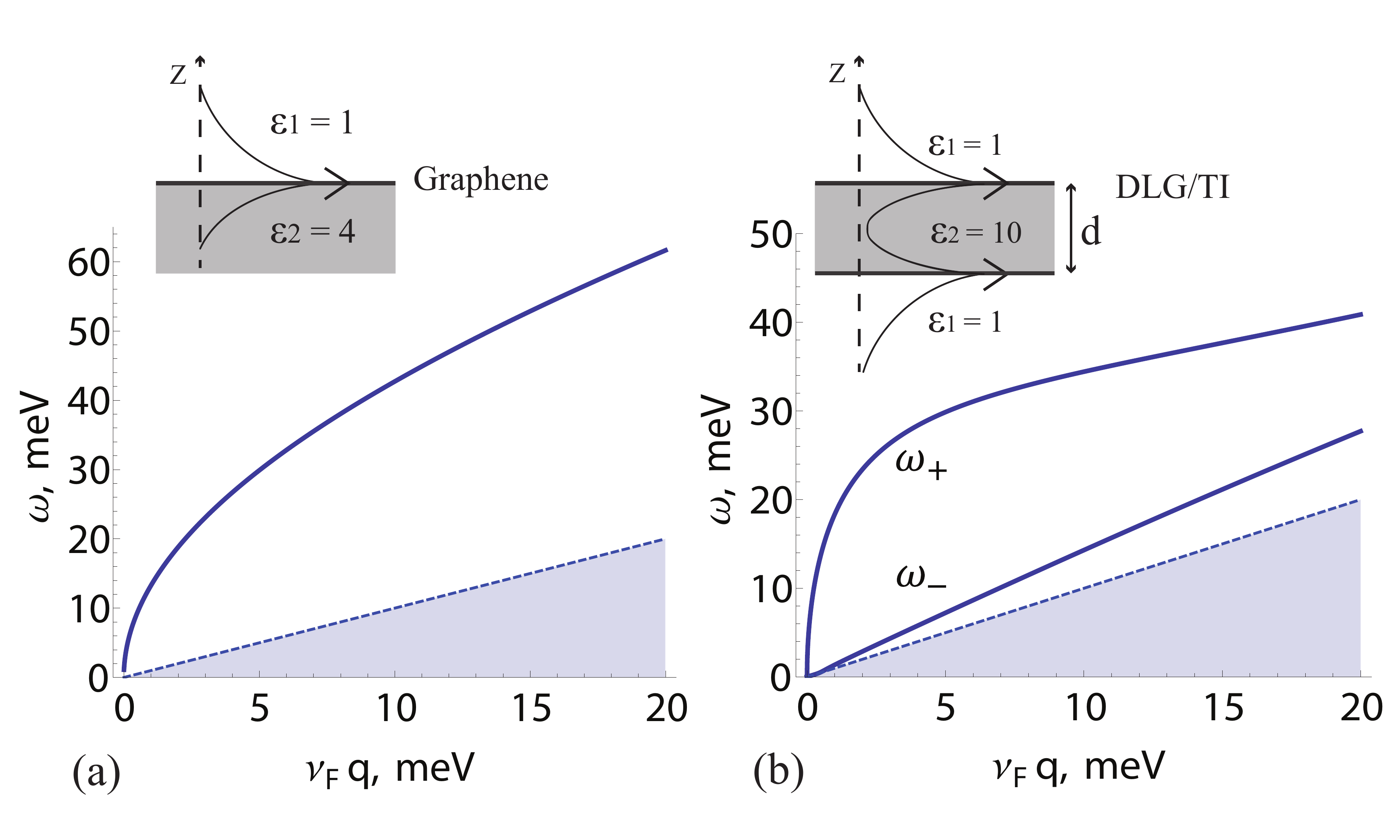}}
  \caption{(a) Dispersion curve of a surface plasmon mode in monolayer graphene  for $E_F = 50 $ meV. The dielectric constant  $\epsilon_2 = 4$ corresponds to SiO$_2$ in the THz region. (b) Dispersion curves for symmetric ($\omega_+$) and antisymmetric ($\omega_-$) plasmon modes in a thin TI film forming a double layer with a spacer thickness $d=6$ nm; $E_F=100$ meV. Electric field of the symmetric plasmon mode is shown in the inset. The value of $\epsilon_2 = 10$ typical for semiconductors was assumed.  Shaded area is the Landau damping region.}
  \end{figure}

Its solution for the real parts of the frequency and wavenumber is shown in Fig. 1(a). When $cq \gg \omega \gg  \upsilon_F q$, the dispersion relation can be simplified to
$ 4\pi\tilde{\chi}_{\omega,q} q + \epsilon_1+\epsilon_2 \approx 0$,
and further to the familiar dependence \cite{kats} $\omega(q) \propto \sqrt{E_F q}$ if we neglect the $q$-dependent term in the denominator of Eq.~(\ref{chi1}).
With increasing plasmon frequency, the plasmon-phonon coupling and interband transitions need to be taken into account \cite{phonon}.

The double layer geometry of the kind shown in the inset to Fig. 1(b) supports two types of surface plasmon modes: symmetric $\omega_+$ and antisymmetric $\omega_-$ (only the field of the symmetric mode is shown in the inset).  Such a geometry appears naturally in thin films of TIs and can be also implemented by separating two graphene layers with a dielectric. The plasmon modes in a double layer were studied theoretically in \cite{TI_plasmon,hwang}.  Here, instead of using the static dispersion relation $\epsilon(\omega,q) = 0$, we start from the full Maxwell's equations and derive the following dispersion equations for the symmetric (top) and antisymmetric (bottom) modes:
\begin{equation}
\label{double}
D(\omega,q)=4\pi\tilde{\chi}_{\omega,q}+\frac{\epsilon_1}{p_1}+\frac{\epsilon_2}{p_2}
\left\{\begin{array}{c}
\tanh(p_2d/2)\\
\coth(p_2d/2)\end{array}
\right\}=0,
\end{equation}
where $d$ is the distance between two 2D layers of massless Dirac electrons. Generalization to the case when the top and bottom media have different dielectric constants is straightforward, but makes the equations more cumbersome. The solution to the dispersion equations is shown in Fig. 1(b).  We consider the limit when $d$ is thicker than about 5 nm  so that the electron hybridization and tunneling can be neglected \cite{zhang}, but on the other hand, $d$ is thin enough to satisfy $p_2d \ll 1$ which ensures strong electromagnetic coupling. For the plasmon frequency of 1 THz the latter means $d \ll 1$ $\mu$m. In the limit of a thick spacer $p_2d \rightarrow \infty$, the $\omega_+$ and $\omega_-$ modes merge and turn into uncoupled monolayer plasmon modes supported by each surface.
\begin{figure}[htb]
\centerline{
\includegraphics[width=9cm]{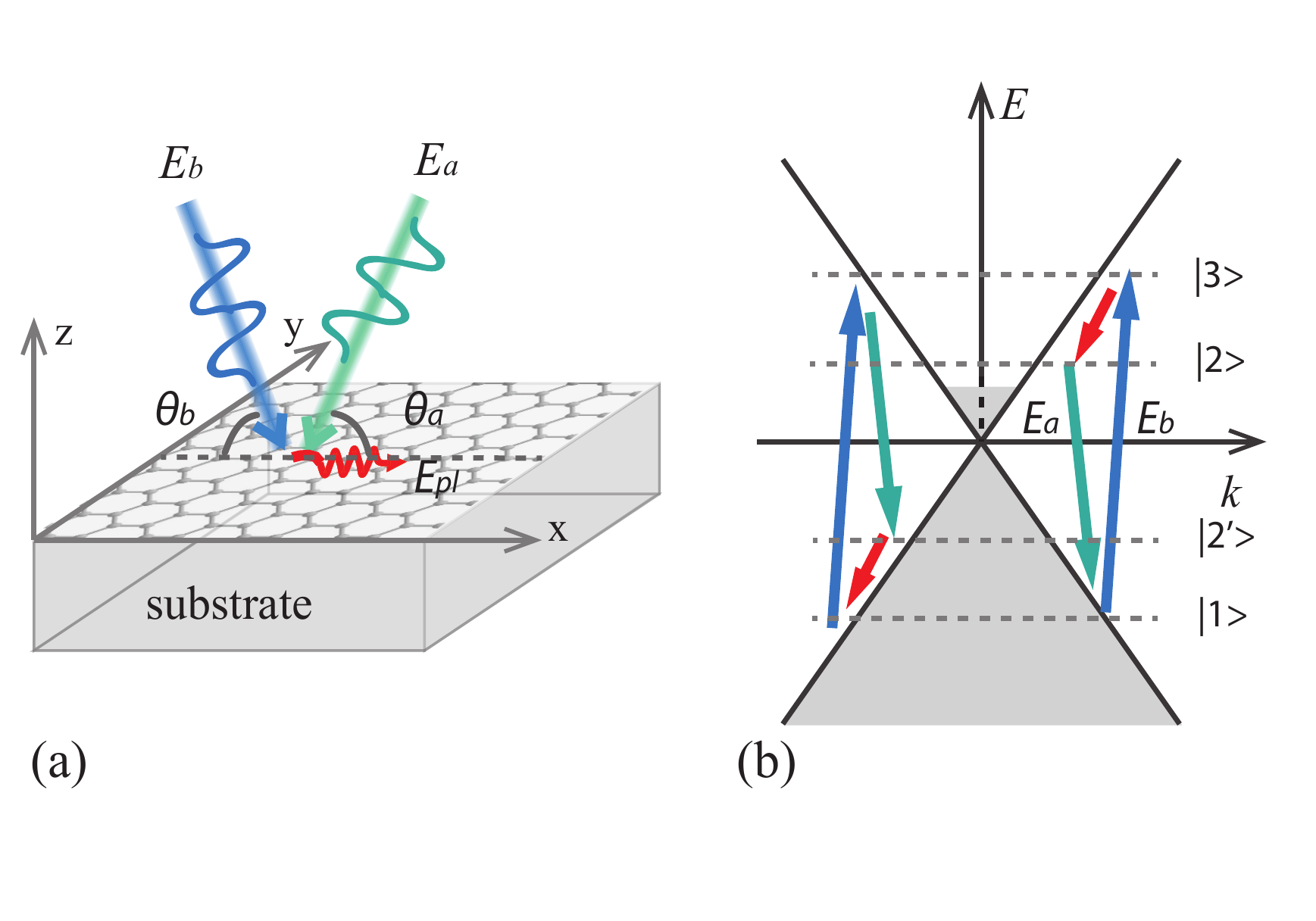}}
  \caption{(a): Geometry of the DFG process. Two pump fields at frequencies $\omega_a$ and $\omega_b$ incident at angles $\theta_a$ and $\theta_b$ on graphene/TI placed on a substrate generate a highly confined surface plasmon field $E_{pl}$ at their difference frequency and in-plane wave vector $q = q_b - q_a$. (b): Elementary three-wave-mixing processes involving two photons and a plasmon coupled to interband and intraband transitions, respectively. Grey shading indicates filled electron states.}
  \end{figure}

Now we turn to the nonlinear optical excitation of THz surface plasmons. First, consider two incident mid-infrared pump fields  at frequencies $\omega_a$ and $\omega_b$ linearly polarized in the $xz$ plane in the geometry shown in Fig. 2(a). For high pump frequencies a purely intraband contribution to the second-order susceptibility is very small, and the three-wave mixing processes that give the main contribution to the DFG signal at frequency $\omega = \omega_b - \omega_a$ are those in which the pump fields are coupled to the interband transitions and the difference-frequency field is coupled to the intraband transitions, as sketched in Fig. 2(b). From the density matrix equations we can derive the 2D second-order nonlinear susceptibility \cite{suppl}:
\begin{eqnarray}
\label{chi2}
&&\chi^{(2)}_{ijk} = \frac{g e^2}{\hbar^2\omega_a\omega_b}\iint \frac{d^2\textbf{k}_1}{(2\pi)^2}\left\{ \left[
\frac{f(\textbf{k}_1)-f(\textbf{k}_3)}{\omega_{31}-\omega_b-i\gamma}\right.\right. \nonumber\\
&&+ \left.\frac{f(\textbf{k}_1)-f(\textbf{k}_2)}{-\omega_{21}+\omega_a-i\gamma}\right]
\frac{\mu^i_{32}v^j_{31}v^k_{12}}{\omega_{32}-\omega-i\gamma} -
\left[\frac{f(\textbf{k}_1)-f(\textbf{k}_3)}{\omega_{31}-\omega_b-i\gamma}\right.\nonumber\\
&&\left.\left.+\frac{f(\textbf{k}_{2'})-f(\textbf{k}_3)}{-\omega_{32'}+\omega_a-i\gamma}\right]
\frac{\mu^i_{2'1}v^j_{31}v^k_{2'3}}{\omega_{2'1}-\omega-i\gamma}\right\},
\end{eqnarray}
where $f(\textbf{k})$ is the occupation number of a given $\textbf{k}$-state and $\textbf{k}_2(\textbf{k}_{2'})$ and $\textbf{k}_3$ can be expressed through $\textbf{k}_1$ utilizing the selection rules that follow from the matrix elements of the interaction Hamiltonian: $\textbf{k}_1 + \textbf{q}_b = \textbf{k}_3$; $\textbf{k}_1(\textbf{k}_{2'}) + \textbf{q}_a = \textbf{k}_2(\textbf{k}_3)$.
Here $\displaystyle \vec{v}_{\alpha \beta} = \upsilon_F\langle \alpha |\vec{\sigma}| \beta\rangle$ is the matrix element of the velocity operator, $\displaystyle \vec{\mu}_{\alpha \beta} = \frac{ie\upsilon_F}{\omega_{\beta \alpha}}\langle \alpha |\vec{\sigma}| \beta \rangle$ is the transition dipole matrix moment, $\vec{\sigma} = (\hat{\sigma}_x,\hat{\sigma}_y)$ is a 2D Pauli matrix-vector, and $\omega_{\alpha \beta} = (\varepsilon(\textbf{k}_{\alpha})-\varepsilon(\textbf{k}_{\beta}))/\hbar$ is the energy difference between electron states $|\alpha \rangle$ and $|\beta \rangle$.

An order of magnitude estimate for $\chi^{(2)}$ can be obtained if we take the limit of a degenerate electron distribution $k_B T \rightarrow 0$ and assume that $\omega \gg \upsilon_F q \cos\phi, \gamma$ and $\omega_a \simeq \omega_b$. Then Eq. (\ref{chi2}) gives
\begin{equation}
\label{chi2app}
\chi_{xxx}^{(2)} \simeq \frac{e^3}{8 \pi \hbar^2} \frac{g}{q \omega_a \omega} \left\{ \frac{\pi}{2} + \arctan\left(\frac{\omega_a - 2 \upsilon_F k_F}{\gamma}\right) \right\},
\end{equation}
where $k_F$ is the Fermi momentum. This approximate expression matches well the low-temperature dependence of the magnitude of   $\chi^{(2)}$ for graphene shown in Fig. 3(a) as a function of the Fermi energy. With increasing Fermi energy, more states in the integral in Eq. (4) are fully occupied, which leads to a lower $\chi^{(2)}$. The value of $|\chi^{(2)}| \sim 10^{-6}$ esu is extremely high: if we divide it by a monolayer thickness $\sim 0.3$ nm to compare
 with bulk susceptibilities of other nonlinear materials, we obtain $|\chi^{(2)}_{bulk}| \sim 10^{-2}$ m/V, which is four orders of magnitude higher than $|\chi^{(2)}|$ measured at similar wavelengths in asymmetric coupled quantum-well structures \cite{sirtori1994,belkin2007}.
\begin{figure}[htb]
\centerline{
\includegraphics[width=7cm]{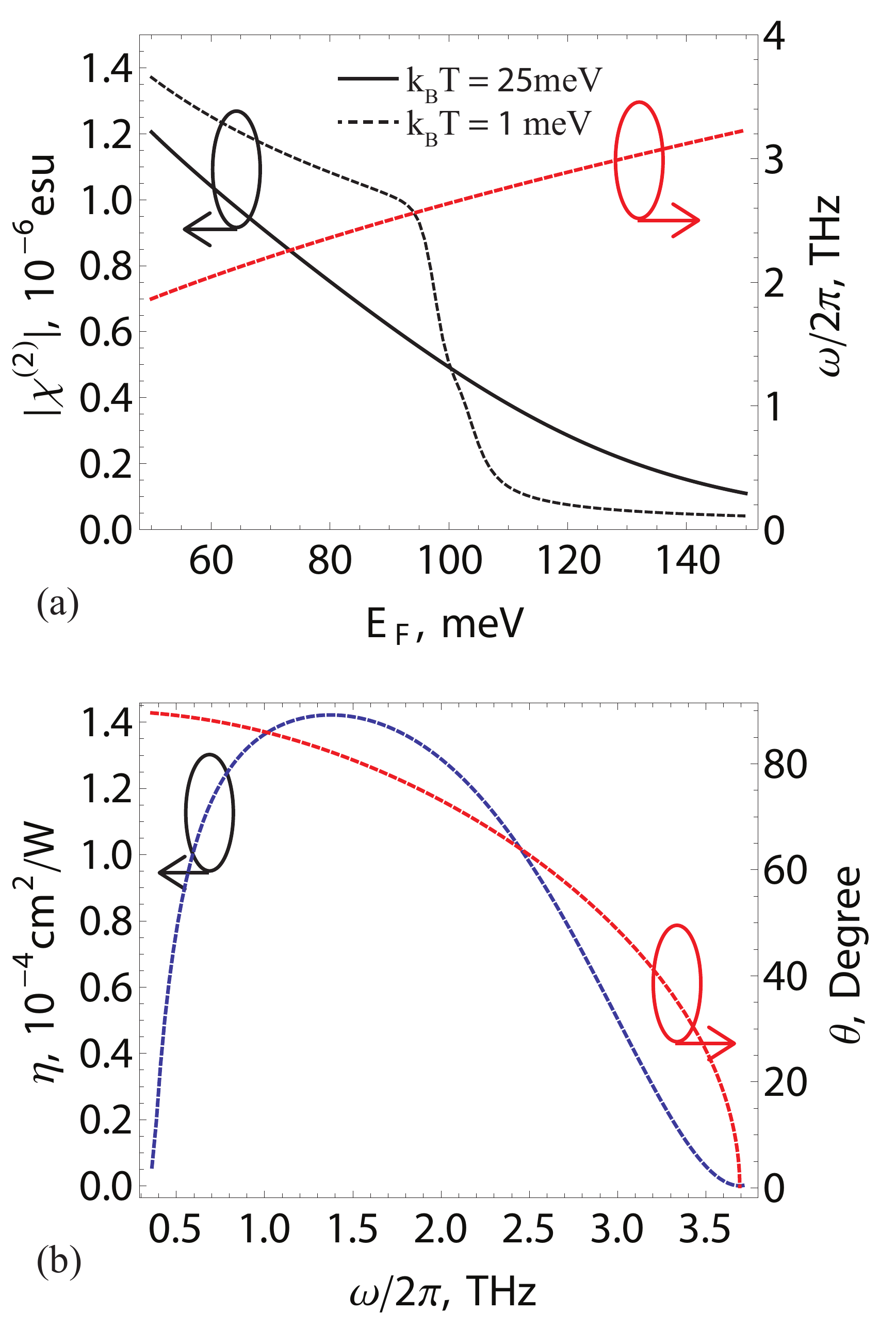}}
  \caption{(a) Magnitude of the 2D $\chi_{xxx}^{(2)}$ as a function of Fermi energy for a fixed incidence angle $\theta_a = \theta_b = \theta = 60^{\circ}$ and fixed sum of the incident pump frequencies $\omega_a + \omega_b = 400$ meV. Red line is the generated plasmon frequency which satisfies the phase-matching condition and energy conservation at each $E_f$.
  (b) The DFG efficiency $\eta = \frac{I_{pl}}{I_a I_b}$ and incidence angle $\theta$ as a function of the plasmon frequency under the conditions of frequency and phase matching. The sum of the incident pump frequencies is fixed to $\omega_a + \omega_b = 400$  meV.  }
  \end{figure}

Note that, unlike the situation in a magnetized graphene with a discrete energy spectrum \cite{yao,tokman}, here we don't have strong resonant enhancement of the nonlinear susceptibility at interband resonances: it is smeared out by integration over the continuous spectrum of electron momenta, as is clear from Eq. (\ref{chi2app}).  Also, the second-order susceptibility is not affected by plasmon resonance; its large value is mainly due to large matrix elements of single-particle transitions. At the same time, the generated difference-frequency field does experience a strong enhancement at the surface plasmon resonance, i.e. when the momentum and frequency of the generated difference-frequency photons satisfy the plasmon dispersion relation given by Eq. (\ref{mono}) or (\ref{double}), depending on the layer geometry.

Indeed, as follows from Maxwell's equations, for a given 2D (surface) polarization $\vec{P} \propto \hat{x}e^{iqx - i\omega t}$ the generated surface plasmon mode has the electric field amplitude \cite{suppl} 
\begin{equation}
\label{res1}
E_{pl}(\omega,q) = -\frac{4\pi}{D(\omega,q)}P =
-\frac{\chi^{(2)}_{xxx}(E_a)_x(E_b)_x}{{\rm Im}{\tilde{\chi}_{\omega,q}}}.
\end{equation}
Here the second equality is valid when $q$ and $\omega$ satisfy the plasmon dispersion equation, in which case the real part of the denominator $D(\omega,q)$ goes to zero and the generated field is greatly enhanced. For example, in monolayer graphene, when $q >> \sqrt{\epsilon_{1,2}}\omega/c$, we obtain
\begin{equation}
\label{res2}
\displaystyle {\rm Im}{\tilde{\chi}(\omega,q)} \approx \frac{\gamma}{4\pi\omega}\frac{\epsilon_1+\epsilon_2}{q}.
\end{equation}

Fig. 3(b) shows the DFG efficiency $\eta$ as a function of the plasmon frequency, where $\eta = \frac{I_{pl}}{I_a I_b}$ is defined as a ratio of the plasmon field intensity to the product of intensities of the incident pump fields.
The generated frequency $\omega$ can be tuned by varying the pump frequencies $\omega_{a,b}$ or/and the angles of incidence $\theta_{a,b}$. The efficiency goes to zero at $\theta_{a,b} = 0$ (because $(E_{a,b})_x \propto \sin{\theta_{a,b}}$) and at $\theta_{a,b} = \pi/2$ when $\chi^{(2)}$ vanishes by symmetry. For mid-infrared pump frequencies, the generated signal can be tuned from 1 THz to several THz while still maintaining a high efficiency; see Fig.3b. For example, focusing two 1-W mid-infrared beams at wavelengths around 5 $\mu$m into the area of $100\times 100$ $\mu$m$^2$ yields about 0.01 W of power in the in-plane propagating THz plasmon mode. Note that the DFG efficiency scales roughly as $1/\gamma^2$ according to Eqs.~(\ref{res1},\ref{res2}). The relaxation rate $\gamma$ is 1 meV for the  plots in Fig. 3.

For integrated photonics and optoelectronics applications, it is desirable to avoid open optical paths and integrate all fields into a planar waveguide structure, see  e.g. \cite{integrated,tredicucci}. The simplest geometry is a dielectric or semiconductor waveguide with graphene deposited at the interface between the core and the cladding.  Fig. 4 shows one such example: a Si/SiO$_2$ waveguide in which the top cladding is air, with graphene deposited on the top. Another example would be a TI film in which the bulk TI material serves as a waveguide core. Consider a waveguide with a core layer of thickness $d$ and dielectric constant $\epsilon_2$ surrounded by cladding materials of lower dielectric constants $\epsilon_1$ and $\epsilon_3$.  Let the two pump fields propagate in opposite directions as fundamental TM modes of the waveguide. They can be excited e.g. by $z$-polarized laser beams. In a TM mode the  longitudinal component of the electric field $E_x \propto \partial B_y/\partial z$ has a maximum near the interfaces $z = \pm d/2$, where $z = 0$ in the middle of the waveguide core.  Strong overlap of the $E_x$ components of the pump fields with a graphene layer is exactly what we need for efficient nonlinear excitation of a surface plasmon at the difference frequency. 

For the $E_x$-fields in the two pump modes given by $E_a \sim \exp{(iq_a x-i\omega_a t)}$ and $E_b \sim \exp{(-iq_b x -i\omega_b t)}$, the second-order nonlinear interaction between the fields generates the 2D nonlinear polarization in graphene at the difference frequency $\omega = \omega_b - \omega_a$ and in-plane momentum $q = q_a + q_b$. When $\omega$ and $q$ satisfy the surface plasmon dispersion relation, the  plasmon field excited by the nonlinear polarization experiences a strong enhancement, similarly to Eq. (6) \cite{suppl}: 
\begin{equation}
E_c(\omega, q) =   -\frac{\chi^{(2)}_{xxx}E_a(\frac{d}{2})E_b(\frac{d}{2})}{{\rm Im}\tilde{\chi}(\omega,q)}.
\end{equation}
Here we assumed that the pump fields are constant over the vertical confinement scale of the plasmon mode $\sim 1$ $\mu$m. 
\begin{figure}[htb]
\centerline{
\includegraphics[width=9cm]{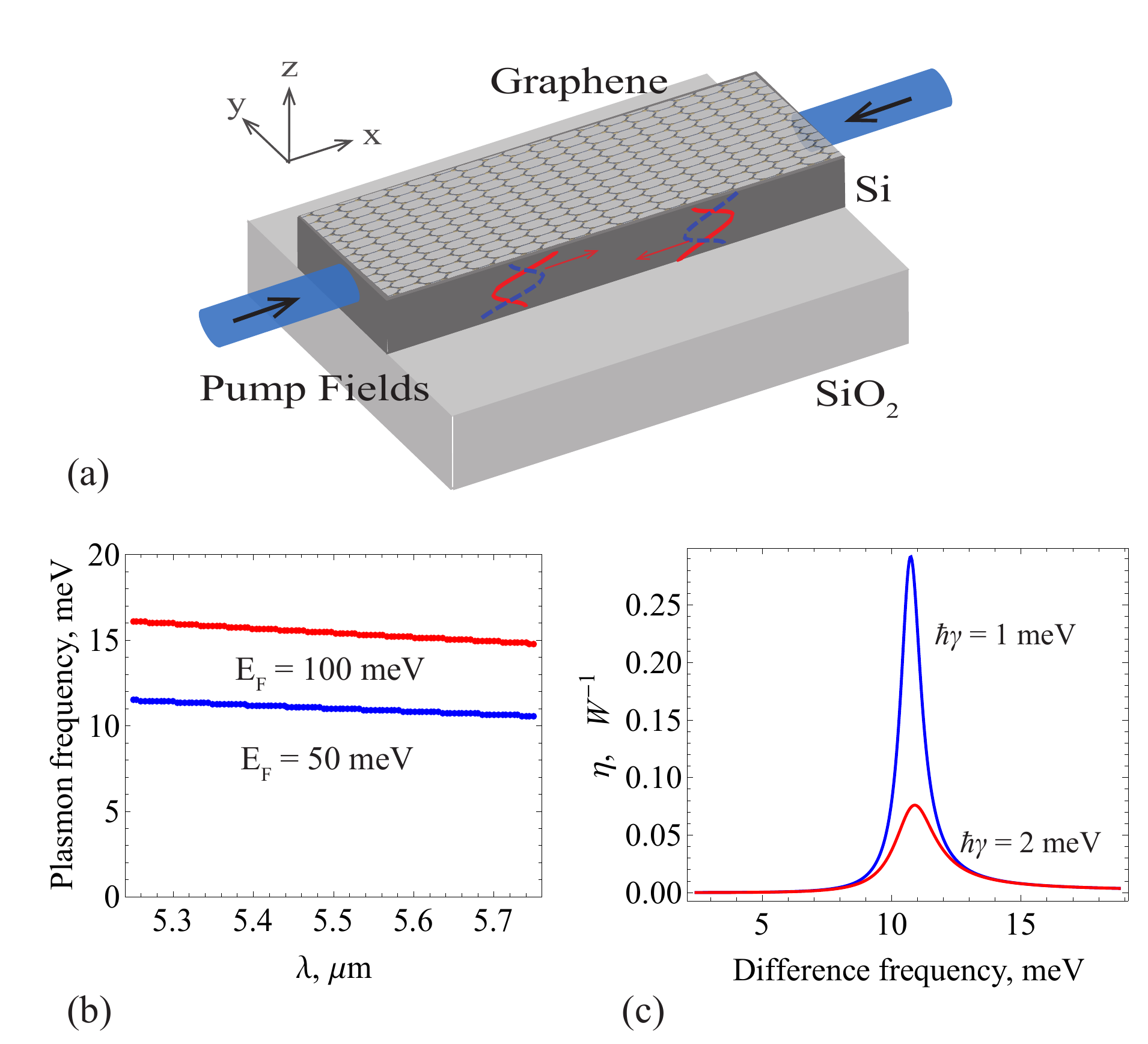}}
\caption{(a) Integrated waveguide geometry of DFG of surface plasmons by counter-propagating TM modes. Profiles of $B_y$ and $E_x$ field components are indicated in blue dashed and red lines, respectively. (b) Phase-matched plasmon frequency vs. wavelength of the fundamental TM mode for one of the pump fields in a 1-$\mu$m thick Si waveguide, for two values of Fermi energy. (c) DFG efficiency of surface plasmons for $E_F = 50$ meV and two values of the relaxation rate.  }
\end{figure}
In Fig.~4 the DFG efficiency is defined as a ratio of the Poynting flux in the surface plasmon mode to the product of Poynting fluxes of the TM modes of the pump fields. Using mid-IR quantum cascade lasers of a typical power 0.1-1 W as a pump, one can get  about 1-10 mW of power in the coherent plasmon mode. We also performed similar calculations for pump lasers at telecom wavelengths around 1.5 $\mu$m. In this case the phase-matched DFG of surface plasmons takes place around plasmon frequency 25-35 meV (for $E_F$ of 50-100 meV) and the conversion efficiency is $\sim 10^{-4}-10^{-5}$/W.  

In conclusion, we found that THz surface plasmons in graphene and topological insulators can be generated with high efficiency through second-order nonlinear frequency mixing of two obliquely incident or in-plane propagating mid-IR beams over a broad range of frequencies and angles of incidence. This process can be used for nonlinear generation, detection, and modulation of the THz light in these fascinating materials or for the manipulation of electron states by optical means.

This work has been supported by NSF Grants OISE-0968405 and EEC-0540832, and by Russian Foundation  for  Basic  Research Grants 13-02-00376 and 13-02-97039.

\end{document}